\documentclass[runningheads]{llncs}
\usepackage{hyperref}
\usepackage[super]{nth}
\usepackage[ruled]{algorithm2e}

\usepackage{amsmath,amsfonts,bm}

\def\eqref#1{equation~\ref{#1}}

\def\1{\bm{1}}

\DeclareMathAlphabet{\mathsfit}{\encodingdefault}{\sfdefault}{m}{sl}
\SetMathAlphabet{\mathsfit}{bold}{\encodingdefault}{\sfdefault}{bx}{n}

\usepackage{float, graphicx, wrapfig, subcaption}
\usepackage{booktabs, tabularx, multirow, tablefootnote}

\usepackage{url}
\usepackage{nicefrac}
\usepackage{comment}
\usepackage{color}
\usepackage{colortbl}
\usepackage{xcolor}
\usepackage{colortbl}
\usepackage{hyphenat}
\usepackage{algorithmic}
\usepackage{enumitem}
\usepackage[justification=centering]{caption}
\usepackage[justification=centering]{subcaption}

\DeclareMathOperator*{\argmaxB}{argmax}   

\begin{document}
\title{HyperNetVec: Fast and Scalable Hierarchical Embedding for Hypergraphs}
%
%
\author{Sepideh Maleki\inst{1} \and
Donya Saless\inst{2} \and
Dennis P. Wall \inst{3} \and
Keshav Pingali\inst{1}
}
\authorrunning{S. Maleki et al.}
%
\institute{The University of Texas at Austin, Austin TX, USA 
\email{\{smaleki,pingali\}@cs.utexas.edu}\and
The University of Tehran , Tehran, Iran \\
\email{donya.saless@ut.ac.ir}\and
Stanford University, Stanford CA, USA \\
\email{dpwall@stanford.edu}}
\maketitle              
\begin{abstract}
Many problems such as node classification and link prediction in network data can be solved using graph embeddings. However, it is difficult to use graphs to capture 
non-binary relations such as communities of nodes. These kinds of 
complex relations are expressed more naturally as hypergraphs.
While hypergraphs are a generalization of graphs, state-of-the-art graph
embedding techniques are not adequate for solving prediction and
classification tasks on large hypergraphs accurately in reasonable time.
In this paper, we introduce HyperNetVec, a novel hierarchical framework for scalable unsupervised hypergraph embedding.
HyperNetVec exploits shared-memory parallelism and is capable of generating high quality embeddings for real-world hypergraphs with millions of nodes and hyperedges in only a couple of minutes while existing hypergraph systems either fail for such large hypergraphs or may take days to produce the embeddings.

\end{abstract}
\section{Introduction}
\label{sec:intro}

A \emph{hypergraph} is a generalization of a graph in which an edge can connect any number of nodes. Formally, a hypergraph $H$ is a tuple $(V,E)$ where $V$ is the set of \emph{nodes} and $E$ is a set of nonempty subsets of $V$ called {\em hyperedges}. Nodes and hyperedges may have weights. Graphs are a special case of hypergraphs in which each hyperedge connects exactly two nodes.

Hypergraphs arise in many application domains. For example, Giurgiu et al. ~\cite{crum} model protein interaction networks as hypergraphs; nodes in the hypergraph represent the proteins and hyperedges represent {\em protein complexes} formed by interactions between multiple proteins. The DisGeNET knowledge  platform \cite{disgenet} represents a disease genomics dataset as a hypergraph in which nodes represents genes and hyperedges represent diseases associated with certain collections of genes. Algorithms for solving hypergraph problems are then used to predict new protein complexes or to predict that a cluster of genes is associated with an as-yet undiscovered disease.

\subsection{Hypergraph Embedding}

Bengio et al. \cite{rl} show that one way to solve prediction problems in graphs is to find an embedding of the graph using {\em representation learning}. Formally, an embedding of a network is a mapping of the vertex set into $\mathcal{R}^d$ where $n$ is the number of nodes in the network and $d << n$. There is a rich literature on graph embedding methods that use a variety of techniques ranging from random walks \cite{deepwalk,line,node2vec} to matrix factorization \cite{fact} and graph neural networks \cite{graphsage,gin}. Graph embedding techniques can be extended to hypergraphs in two ways but neither of them is satisfactory.

One approach is to represent the hypergraph as a graph by replacing each hyperedge with a clique of edges connecting the vertices of that hyperedge, and then use graph embeddings to solve the prediction problems. This approach has been explored in HGNN \cite{hgnn} and HyperGCN \cite{hypergcn}. However, the clique expansion is lossy because the hypergraph cannot be recovered from the clique expansion in general. This information loss persists even if the dual of the hypergraph is considered~\cite{lossy}.

Zien et al. \cite{star} show that another approach is to work with the \textit{star expansion} of the hypergraph. 
Given a hypergraph $H {=} {(}V{,}E {) }$ where $V$ is the set of nodes and $E$ is the set of hyperedges, we create a bipartite graph  $H^{\ast} = (V^{\ast},E^{\ast})$ by (i) introducing a node $v_e$ for each hyperedge $e \in E$ so in final graph $V^{\ast} = V \cup E$, and (ii) introducing an edge between a node $u \in V$ and a hyperedge node $v_e \in E$ if $u \in e$ in the hypergraph, so 
{\em i.e.,} $E^{\ast} = {(u,v_e) : u \in e, e \in E}$. Unlike the clique expansion of a hypergraph, the star expansion is not lossy provided nodes representing hyperedges are distinguished from nodes representing hypergraph nodes. However, graph representation learning approaches do not distinguish between the two types of nodes in the bipartite graph, which lowers accuracy for prediction problems as we show in this paper.

These problems motivated us to develop \textit{HyperNetVec}, a {\em parallel multi-level framework for constructing hypergraph embeddings}. HyperNetVec leverages existing graph embedding algorithms and it performs hypergraph embedding in a much faster and more scalable manner than current methods. We evaluate HyperNetVec on a number of data sets for node classification and hyperedge prediction. Our experiments show that our hierarchical framework can compute the embedding of hypergraphs with millions of nodes and hyperedges in just a few minutes without loss of accuracy in downstream tasks, while all existing hypergraph embedding techniques either fail to run on such large inputs and or take days to complete. 
 
\begin{figure}[ht]
\centerline{\includegraphics[width=0.55\columnwidth,page=3]{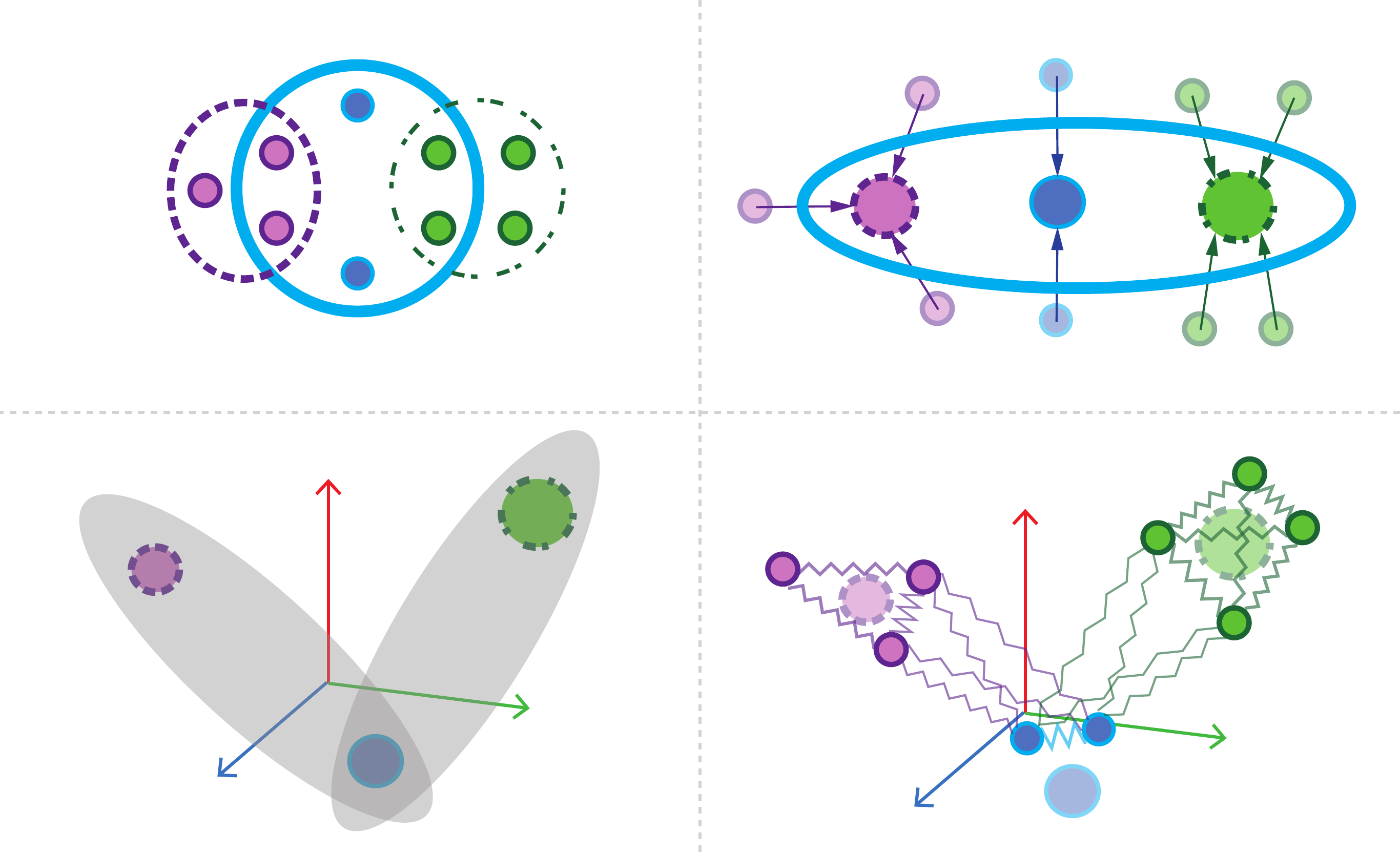}}
\caption{Multi-level embedding. \textbf{Top left}: original hypergraph. \textbf{Top right}: contracting nodes of a hypergraph to create a coarser hypergraph. \textbf{Bottom left}: initial embedding. \textbf{Bottom right}:  iterative refinement of embeddings.}
\label{fig:multilevel}
\vskip -0.2in
\end{figure}

Our main technical contributions are summarized below.
\begin{itemize}
\item \textbf{HyperNetVec:} We describe \emph{HyperNetVec}, a hierarchical hypergraph embedding framework that is designed to handle variable-sized hyperedges.

\item \textbf{Node features:} Unlike many other systems in this space, \textbf{HyperNetVec} can exploit the topology of the graph as well as node features, if they are present in the input data. 

\item \textbf{Unsupervised hypergraph embedding system:} To the best of our knowledge, HyperNetVec is the first unsupervised hypergraph embedding system. The embeddings obtained from this framework can be used in downstream tasks such as hyperedge prediction and node classification.

\item \textbf{Scalability:} HyperNetVec is the first hypergraph embedding approach that can generate embeddings of hypergraphs with millions of nodes and hyperedges.

Our approach can significantly reduce run time while producing comparable and in some cases, better accuracy than state-of-the-art techniques. 

\end{itemize}

\section{Related Work}


There is a large body of work on graph and hypergraph embedding techniques so we discuss only the most closely related work.

\subsection{Network Embedding}
\label{sec:emb}

There are relatively few efforts on hypergraph embedding that treat hyperedges as first-class entities. As mentioned above, one popular approach to hypergraph embedding is to convert the hypergraph to a graph and then use a graph embedding technique\cite{hypergraph2,hypergraph3,hgnn}. For example, each hyperedge can be replaced with a clique connecting the nodes of that hyperedge to produce a graph representation. Other approaches such as HyperGCN ~\cite{hypergcn} use a graph convolution on a modified clique expansion technique where they \textit{choose} what edges to keep in the graph representation. 
While this method keeps more structure than methods based on the clique expansion of a hypergraph, existing methods fail to scale to large networks as we show in this paper.

\subsubsection{Multi-level Embedding}
Multi-level (hierarchical) approaches attempt to improve the run-time and quality of existing or new embedding techniques. 
Multi-level graph embedding consists of three phases: coarsening, initial embedding, and refinement.
~\textit{Coarsening}: A coarsened graph $G^{\prime}$ is created by merging pairs of nodes in input graph $G$. This process is applied recursively to the coarser graph, creating a sequence of graphs. in which the final graph is the coarsest graph that meets some termination criterion ({\em e.g.}, its size is below some threshold). 
~\textit{Initial embedding}: Any unsupervised embedding methods for networks can be used to generate an initial embedding. 
~\textit{Refinement}: For graphs $G^{\prime}$ and $G$, the embedding of $G^{\prime}$ is projected onto $G$ and then refined,
starting from the coarsest graph and finishing with the original graph.

There is a large body of research on multi-level \textit{graph} embedding. For example, \textit{HARP} \cite{harp} generates a hierarchy of coarsened graphs and perform embedding from the coarsest level to the original one. MILE \cite{mile} uses heavy edge matching \cite{metis1} to coarsen the graph and leverages GCN as a refinement method to improve embedding quality. However, training a GCN model is very time consuming for large graphs and leads to poor performance when multiple GCN layers are stacked together \cite{deeper}. These multi-level embedding methods only utilise structural information (topology) of a graph. However, in many datasets such as citations, nodes of a graph have attributes. For a high quality embedding, it is important to exploit node attributes as well as structural information of a graph. GraphZoom\cite{graphzoom} first performs graph fusion to generate a new graph that encodes the topology of the original graph and the node attribute information and then uses a coarsening algorithm that merges nodes with high spectral similarity. Finally, they apply a local refinement algorithm. While GraphZoom outperforms previous multi-level embedding systems, it still takes hours, in some cases days, to generate embeddings for a graph with millions of nodes and edges. 

In general, hypergraphs are a more complicated topic and the corresponding algorithms are typically more compute and memory intensive. Multi-level approaches for \textit{hypergraphs} have been used mainly for hypergraph partitioning \cite{hmetis,phmetis,bipart}. In principle, ideas from multi-level graph embedding approaches can be adopted for hypergraphs. 
For example, for the coarsening algorithm, we can  merge pairs of nodes that have a hyperedge in common (heavy edge matching). While this approach is able to produce coarser hypergraphs, it reduces the number of hyperedges in the coarser graphs only
for those pairs of matched nodes that are connected by a hyperedge of size two. As a result, the coarsest hypergraph is still large in terms of the size of the hyperedges which increases the running time of the overall algorithm. In our experience, the coarsening and refinement algorithms proposed in multi-level graph embedding systems are not adequate for solving inference problems on hypergraphs, as we discuss in this paper. These limitations led us to design HyperNetVec, which scales to hypergraphs with millions of nodes and hyperedges while producing high-quality embeddings. 
\section{Methodology}
\label{sec:methodology}

Given a hypergraph $H {=} {(}V{,}E{) }$, the algorithms described in this paper use the star expansion of the hypergraph and assign a vector representation $h_u$ to each $u {\in} {(}V {\cup} E{)}$. Intuitively, these embeddings attempt to preserve {\em structural similarity} in the hypergraph: if two hyperedges have many nodes in common or if two nodes are in many of the same hyperedges, the algorithm attempts to assign the two hyperedges/nodes to points that are \textit{close} in the vector space. \textit{Closeness} can be computed using distance or other measures of vector similarity. Embedding should also exploit the \textit{transitivity} property of similarity: if $a$ and $b$ are similar, and $b$ and $c$ are similar, we want the embedding of $a$ and $c$ to be close to each other as well. Finally, if nodes have features, the embeddings should also exploit {\em functional similarity} between nodes. 

Figure~\ref{fig:multilevel} illustrates the high-level idea of  multi-level hypergraph embedding.  This framework consists of three phases: (i) {\em Coarsening}, which iteratively merges nodes of the hypergraph to shrink the size of the hypergraph until the hypergraph is small enough that any network embedding algorithm can quickly obtain the embedding of the smallest hypergraph; (ii) {\em Initial embedding}, in which a network embedding algorithm is used on the coarsest hypergraph to generate the embedding, and (iii) {\em Refinement}, in which the embedding vectors of the coarser hypergraph are projected onto a finer hypergraph and a refinement algorithm is used to refine these embedding vectors. In the rest of this section, we describe these phases in more detail. HyperNetVec is a parallel implementation of the multilevel approach.   

\subsection{Coarsening}
\label{sec:coarsening}

Intuitively, coarsening finds nodes that are \textit{similar} to each other and merges them to obtain a coarser hypergraph. To obtain a high quality embedding, we need to explore both structural similarity and functional similarity. The connectivity of nodes and hyperedges of a hypergraph determines structural similarity while node features determine functional similarity. 

The first step in coarsening a hypergraph is to find nodes that are similar to each other and merge them. This is accomplished by ``assigning'' each node to one of its hyperedges, and then merging all nodes $\{n_1,n_2,...,n_k\}$ assigned to a given hyperedge to produce a node $n^\prime$ of the coarser hypergraph. We refer to $n^\prime$ as the {\em representative} of node $n_i$ in the coarse hypergraph, and denote it as rep($n_i$). If all nodes of a hyperedge $h_j$ are merged, we remove that hyperedge from the hypergraph. Otherwise, we add the hyperedge to the next level and refer to it as rep($h_j$). If a node $n_i$ is contained in hyperedge $h_j$ in the finer hypergraph and $h_j$ is present in the coarse hypergraph, then rep($n_i$) is made a member of rep($h_j$).

If nodes of a hypergraph have features, this information can be used to find similar nodes and merge them together. 
In a hypergraph with node features, the feature vector of a hyperedge is the mean aggregation of the features of its nodes. In this scenario, metrics of vector similarity or distance between a feature vector of a hyperedge and a node can be used for assigning nodes to hyperedges. However, if the hypergraph has no features, HyperNetVec can use other measures such as weights or degrees of a hyperedge to assign nodes to hyperedges. The datasets used in experiments in this paper for node classification are citation networks and the node feature vectors come from bag-of-words encoding. In these datasets, \textit{cosine similarity} aligns well with class labels. However, other metrics of vector similarity or distance such as L2 norm,  correlation distance, etc. can also be used. 

In summary, at each level of coarsening, HyperNetVec computes a feature vector for a hyperedge by finding the mean aggregation of the feature vectors of its nodes. Then it assigns each node $v$ in the current hypergraph to a hyperedge $c(v)$, defined (for cosine similarity) as $ c(v) = \argmaxB_{e \in \mathcal{N}(v)}~ \dfrac{f(e) \cdot f(v)}{\vert f(e) \vert \cdot \vert f(v) \vert}$
where $\mathcal{N}(v)$ is the set of hyperedges that node $v$ belongs to, $f(e)$ is the feature vector of hyperedge $e$ and $f(v)$ is the feature vector of node $v$.

Nodes that are assigned to the same hyperedge are merged together and the resulting node is added to the coarse hypergraph. In case of a tie, HyperNetVec randomly chooses a hyperedge in the neighborhood of the node. 
\begin{algorithm}[tbh]
   \caption{$\mathsf{Coarsening}$}
   \label{alg:coarse}
\begin{algorithmic}[1] 
   \STATE {\bfseries Input:} $\text{fineGraph}$, node feature matrix $X \in \mathcal{R}^{V \times k}$,\text{neighborhood function} $\mathcal{N}(u)$, \text{depth} $K$
   \STATE {\bfseries Output:} $\text{coarseGraph}$, node feature matrix $X^{\prime} \in \mathcal{R}^{V^{\prime} \times k}$
   
   \FOR{$i=1$ to $K$}

    \FORALL{hyperedges $e \in  fineGraph$ \textbf{in parallel} }
       \STATE {$\mathsf{ComputeEdgeFeature}(e)$} 
     \ENDFOR

     \FORALL{nodes $v \in fineGraph$ in parallel} 
       \STATE {$\mathsf{AssignHyperedge}(v)$} 
     \ENDFOR

     \FORALL{hyperedges $e \in  fineGraph$ \textbf{in parallel} }
       \STATE {$M \leftarrow \mathsf{FindAssignedNodes}(e)$}
       \STATE {$m \leftarrow \mathsf{Merge}(M)$}
       \STATE {$\mathsf{coarseGraph.addNode}(m)$} \# m is representative of each node in M
       \IF{ $|M| \not= |e|$}
           \STATE {$\mathsf{coarseGraph.addHedge}(e)$} \# representative of e
       \ENDIF
     \ENDFOR
      \FORALL{hyperedges $e \in  fineGraph$ \textbf{in parallel} }
      \IF {rep($e$) exists}
      \FOR{$v$ $\in$ $\mathcal{N}(e)$}
        \STATE {$\mathsf{coarseGraph.addEdge}$(rep($e$),rep($v$))}
       \ENDFOR
          \ENDIF
     \ENDFOR
   \ENDFOR
   
\end{algorithmic}
\end{algorithm}
\subsection{Initial Embedding}
\label{sec:initial}
We coarsen the hypergraph until it is small enough that {\em any} unsupervised embedding method can generate the embedding of the coarsest hypergraph in just a few seconds. We use the edgelist of the coarsest bipartite graph (star expansion) as the input to this embedding method.

\subsection{Refinement}
\label{sec:refinement}

The goal of this phase is to improve embeddings by performing a variation of Laplacian smoothing \cite{laplacian} that we call the {\em refinement} algorithm. The basic idea is to update the embedding of each node $u$ using a weighted average of its own embedding and the embeddings of its immediate neighbors $\mathcal{N}(u)$. Intuitively, smoothing eliminates high-frequency noise in the embeddings and tries to assign similar embeddings to nodes that are close to each other in the graph, which improves the accuracy of downstream inference tasks. A simple iterative scheme for smoothing is:
 $\Tilde{z}_{u}^{i} = \sum_{v \in \mathcal{N}(u)} ( \frac{w_{uv}}{\sum_{v {\in}\mathcal{N}(u)}w_{uv}} )z_v^{i-1}$. In this formula, $z_u^i$ is the embedding of node $u$ in iteration $i$, and $w_{uv}$ is the weight on the outgoing edge from $u$ to $v$; if there no weights in the input hypergraph, a value of 1 is used and the denominator is the degree of node $u$. This iterative scheme can be improved by introducing a hyper-parameter $\omega$ that determines the relative importance of the embeddings of the neighboring nodes versus the embedding of the node itself, to obtain the following iterative scheme: $z_{u}^{i} = (1-\omega)z^{i-1}_u + \omega\Tilde{z}_{u}^{i}$. The initial embeddings for the iterative scheme are generated as follows. For the coarsest graph, they are generated as described in section~\ref{sec:initial}. For the other hypergraphs, if a set of nodes $S$ in hypergraph $H_{i{-}1}$ was merged to form a node $n$ in the coarser hypergraph $H_i$, the embedding of $n$ in $H_i$ is assigned to all the nodes of $S$ in $H_{i{-}1}$.

Abstractly, this iterative scheme uses \textit{successive over-relaxation} (SOR) with a parameter $\omega$ to solve the linear system $Lz = 0$ where $L$ is the Laplacian matrix of $H^\ast$, the bipartite (star) representation of the hypergraph. The Laplacian is defined as ($D{-}A$) where $D$ is the diagonal matrix with diagonal elements $d_{uu}$ equal to the degree of node $u$ for unweighted graphs (for weighted graphs, the sum of weights of outgoing edges), and $A$ is the adjacency matrix of $H^\ast$. 
To avoid oversmoothing, we do not compute the exact solution of this linear system but if we start with a good initial embedding $z^{0}$, a few iterations of the iterative scheme lead to significant gains in the quality of the embedding, as we show experimentally in Section~\ref{sec:experiment}. 


\begin{algorithm}[tb]
   \caption{$\mathsf{Refinement}$}
   \label{alg:refinement}
\begin{algorithmic}[1]
   \STATE {\bfseries Input:} \text{Bipartite graph representation} $H^{\ast}=(V^{\ast},E^{\ast},W)$ \text{of hypergraph} $H = (V,E,W)$, \text{vector representation} $z_u$ \text{for all} $u$ $\in$ $(V^{\ast})$, \text{neighborhood function} $\mathcal{N}(u)$, \text{parameter} $\omega$, parameter $k$ for max iteration
   \STATE {\bfseries Output:} \text{Refined vector representation} $h_u$, $\forall{u}$ $\in$ $(V^{\ast})$
   \STATE {$z^0_u \leftarrow z_u,\forall{u}\in (V^{\ast})$}
   \STATE {$iter=0$}
   \WHILE{$iter < k$}
     \FOR{$u \in V^*$ in parallel}
       \STATE {$\Tilde{z}_{u}^{i} \leftarrow \sum_{v{\in}\mathcal{N}(u)} w_{uv}z^{i-1}_v / \sum_{v{\in}\mathcal{N}(u)}w_{uv}$}
       \STATE {$z_{u}^{i} \leftarrow (1-\omega)z^{i-1}_u + \omega\Tilde{z}_{u}^{i}$}
     \ENDFOR
     \STATE{$iter$ += $1$}

   \ENDWHILE
   \STATE {$h_u \leftarrow z^k_u,\forall{u}\in (V^{\ast})$}
   
\end{algorithmic}
\end{algorithm}

Algorithm~\ref{alg:refinement} shows the psuedocode for refinement. The inputs to this algorithm are $H^*$, the bipartite representation of the hypergraph, $z_u$, the initial embedding for each node and hyperedge, and a relaxation parameter $\omega$ between 0 and 1. Embeddings of the hyperedges are updated using the embeddings of the nodes, and the embeddings of nodes are updated using the embeddings of hyperedges. Note that if $u$ represents a hyperedge, $\mathcal{N}(u)$ is the set of nodes in that hyperedge, and if $u$ represents a node in the hypergraph, $\mathcal{N}(u)$ represents the set of hyperedges that $u$ is contained in. Each iteration of the refinement algorithm has a linear time complexity in the size of the bipartite representation of the hypergraph.  

\section{Experiments}
\label{sec:experiment}



HyperNetVec provides an unsupervised method for representation learning for hypergraphs. We show these representations perform well for both node classification and hyperedge prediction. Prior works such as HyperGCN and Hyper-SAGNN have been evaluated for one or the other of these tasks but not both. 

\textbf{Experimental settings.} We implement \textit{HyperNetVec} in Galois 6.0~\cite{pingali}. All experiments are done on a machine running CentOS 7 with 4 sockets of 14-core Intel Xeon Gold 5120 CPUs at 2.2 GHz, and 187 GB of RAM. All the methods used in this study are parallel implementations and we use the maximum number of cores available on the machine to run the experiments. The embedding dimension is 128. The hyperparameter $\omega$ in the refinement algorithm is set to 0.5 for all experiments. Once node embeddings are obtained, we apply logistic regression with cross-entropy loss for our downstream tasks. 

\begin{table}[htbp]
\caption{Datasets used for node classification.}
\label{nodeclass}
\vskip -0.3in
\begin{center}
\begin{tabular}{lcccccr}
\toprule
Dataset & Nodes & Hyperedges & Edges & Classes & Features\\
\midrule

Citeseer & 1,458& 1,079&  6,906&6&3,703 \\
PubMed    &3,840 & 7,963 &69,258 &3&500\\
DBLP    & 41,302& 22,363& 199,122& 6&1,425\\

\bottomrule
\end{tabular}
\end{center}
\vskip -0.3in
\end{table}
\subsection{Node Classification}

Given a hypergraph and node labels on a small subset of nodes, the task is
to predict labels on the remaining nodes. 
We used the standard hypergraph datasets from prior works, and these are listed in Table~\ref{nodeclass}. We are given 4\% of node labels and predict the remaining 96\%. 

{\bf Methods Compared.} We explore a number of popular methods for graph embedding. We also compare our results with hypergraph convolutional networks approaches for semi-supervised classification.

\textit{Random-walk methods:} We select node2vec~\cite{node2vec} (high performance implementation \cite{fastnv}) for this group. This method is properly tuned. We explored window size \{10,20\}, walk length \{20, 40, 80, 120\}, number of walks per vertex \{10,80,40\}, p \{1,4,0.5\}, and q \{1,4,0.5\}. The results reported in the paper are for the best hyper-parameter values, which are 10,80,10,4,1 respectively. 

\textit{Graph convolutional network:} We compare with GraphSAGE~\cite{graphsage}. We use GraphSAGE in unsupervised manner with the mean aggregator model.

\textit{Multi-level based embedding methods:} We compare against unsupervised approaches MILE~\cite{mile}, and GraphZoom~\cite{graphzoom}. MILE is a multi-level graph embedding framework. We used the default refinement technique, MD-gcn. GraphZoom is also a multi-level graph embedding framework. For the coarsening, we used \textit{simple}. 

\textit{Semi-supervised classification on hypergraphs.} We compare with HyperGCN. Given a hypergraph, HyperGCN~\cite{hypergcn} approximates the hypergraph by a graph where each hyperedge is approximated by a subgraph. A graph convolutional network (GCN) is then run on the resulting graph. We used 200 epochs and learning rate of 0.01.
For the multi-level approaches, we use node2vec, and GraphSAGE as the initial embedding methods. Since MILE cannot utilise node features, we do not run GraphSAGE as an initial embedding method for MILE. 
We report the mean test accuracy and standard deviation over 100 different train-test splits. We optimize hyperparameters of all the baselines. For HyperNetVec, we use 80 iterations of refinement. Alternatively, other stopping criteria such as epsilon difference between two consecutive iterations can be used.

\textbf{Running time.} For HyperNetVec and other multi-level approaches, running time includes all three phases: coarsening, initial embedding, and refinement. For the rest of the baselines, we use the time for hypergraph embedding. For each approach to computing the initial embeddings (node2vec, GraphSAGE), we have a row showing the accuracy and running times when that approach is used, and rows below those showing the accuracy and running times if that approach is used in conjunction with HyperNetVec or other multilevel approaches. For example, the first row in Table~\ref{tb:small} shows the running times and accuracy when node2vec is used on the star expansion of the hypergraph, while the second row (HyperNetVec + nv \textit{(l=0)}) shows the total running time and accuracy if HyperNetVec is used without coarsening but with the output of node2vec being post-processed using our refinement algorithm. The line below that \textit{(l=2)} shows the results if two levels of coarsening are used in addition.

{\bf Datasets.} We used the following standard hypergraph datasets in our study.  Nodes not connected to any hyperedge, as well as hyperedges containing only one node, were removed. \textit{Citeseer (co-citation)}: scientific publications classified into six classes. All documents cited by a document are connected by a hyperedge. ~\cite{cora}. \textit{PubMed (co-citation)}: scientific publications from PubMed Diabetes database. All documents cited by a document are connected by a hyperedge~\cite{cora}.
\textit{DBLP (co-authorship)}: scientific publications consist of 6 conference categories. All documents co-authored by an author are in one hyperedge~\cite{dblp}. 

\begin{table*}[htbp]
\caption{Node classification. Accuracy in \% and time in seconds. $l$ is the number of coarsening levels. $0$ means without coarsening. }
\label{tb:small}
\vskip -0.9in
\begin{center}
\begin{tabular}{lcccccc}
\toprule
\textbf{} &   \multicolumn{2}{c}{\textbf{Citeseer}}& \multicolumn{2}{c}{\textbf{PubMed}}& \multicolumn{2}{c}{\textbf{DBLP}} \\ 
  &

   \multicolumn{1}{l}{Accuracy} &
   \multicolumn{1}{l}{Time} &
  \multicolumn{1}{l}{Accuracy} &
   \multicolumn{1}{l}{Time} &
  \multicolumn{1}{l}{Accuracy} &
  \multicolumn{1}{l}{Time} \\

\midrule

node2vec (nv) & 51.3 $\pm 1.$ & 14&$65.3 \pm2.$  & 66 & $64.3 \pm .4$  & 470 \\
HyperNetVec + nv ($l$=$0$)& $59.1$ $\pm 1.$&16&$79.7$ $\pm 1.$ & 70 &$72.4\pm .4$&490\\
HyperNetVec + nv ($l$=$2$)&\textbf{\colorbox{blue!20}{60.6 $\pm 1.$}} & 17&80.7 $\pm 1.$ &69&78.9 $\pm .5$ &    216  \\
GraphZoom + nv ($l$=$2$) & 54.4 $\pm 1.$& 15 & $74.9 \pm .1$ & 100&$70.2 \pm .5$& 434\\
MILE + nv ($l$=$2$)  &  52.2   $\pm 1.$ & 14& $68.7 \pm .2 $& 60 &71.8 $\pm 1.$& 402\\
 \hline
GraphSAGE (gs) & 45.6 $\pm 1.$ &1,167& $60.7 \pm .2$  &    277  &  $67.7 \pm .1$  &  925\\
HyperNetVec + gs ($l$=$0$)&  60.3 $\pm 3.$ &1,170&   $80.4$  $\pm .1$ & 296    &\textbf{\colorbox{blue!20}{79.9 $\pm .1$}}&1,055\\
HyperNetVec + gs ($l$=$2$)& 60.2 $\pm 1.$ & 843& \textbf{\colorbox{blue!20}{80.8 $\pm .1$}}  & 120    &79.4 $\pm .4$ &    530  \\
GraphZoom + gs ($l$=$2$) & 52.7 $\pm 1.$&853&$72.4 \pm .1$  &     137      &  $75.6 \pm .4$&  593\\
 \hline
HyperGCN & 54.1 $\pm 10$  &  12 &$64.3 \pm 10$&60 &$63.3 \pm 10$& 480 \\

\bottomrule
\end{tabular}
\end{center}
\vskip -0.32in

\end{table*}
These are the main takeaways from Table~\ref{tb:small}. HyperNetVec generates the highest quality embeddings for the node classification task. HyperNetVec outperforms HyperGCN in terms of quality for all datasets by up to 15\%. The refinement algorithm improves the quality of embeddings for all the datasets by up to 23\%. 
 This can be seen by comparing the statistics for HyperNetVec without coarsening ($l=0$) with those for node2vec, and GraphSAGE. The initial embedding for HyperNetVec is obtained from node2vec, and GraphSAGE so differences in the statistics arise entirely from the fact that HyperNetVec performs refinement. HyperNetVec outperforms prior multi-level graph embedding approaches (MILE and GraphZoom) for all the datasets by up to 11\% for MILE and up to 9\% for GraphZoom. Coarsening reduces the overall running time of the embedding for larger hypergraphs. Since coarsening reduces the size of the hypergraph, the initial embedding and refinement can be done faster. This can be seen by comparing the statistics for HyperNetVec with 2 levels of coarsening ($l=2$) with those for slower initial embedding approach such as GraphSAGE.

\subsection{Hyperedge Prediction}

In hyperedge prediction, we are given a hypergraph with a certain fraction of hyperedges removed, and given a proposed hyperedge (i.e. a set of nodes) our goal is to predict if this is likely to be a hyperedge or not. Formally, given a $k$-tuple of nodes $(v_1, v_2, ..., v_k)$, our goal is to predict if this tuple is likely to be a hyperedge or not.

We compare our method with the \textit{supervised hyperedge prediction method} Hyper-SAGNN \cite{hypersagnn} on four datasets listed in Table~\ref{tb:link}, and with the graph method node2vec. Hyper-SAGNN is a  self-attention based approach for hyperedge prediction. We used their encoder-based approach with learning rate of 0.001 and 300 epochs.
\begin{table*}[htbp]
\caption{Datasets used for hyperedge prediction.}
\label{tb:link}
\vskip -0.8in
\begin{center}
\begin{small}
\begin{sc}
\begin{tabular}{lccc}
\toprule
\textbf{}   &
  \multicolumn{1}{c}{\textbf{Nodes}} &  \multicolumn{1}{c}{\textbf{Hyperedges}} & \multicolumn{1}{c}{\textbf{Edges}} \\ 
\midrule
\textbf{GPS}   &  221 & 437&1,436 \\
\textbf{MovieLens}   &  17,100&46,413 &47,957  \\
\textbf{Drug}   &   7,486& 171,757  & 171,756 \\
\textbf{Wordnet}   &  81,073& 146,433 & 145,966\\
\textbf{Friendster}   &  7,458,099& 1,616,918 & 37,783,346\\
\bottomrule
\end{tabular}
\end{sc}
\end{small}
\end{center}
\vskip -0.3in
\end{table*}

{\bf Datasets.} We used the following datasets in our study.
\textit{GPS}: a GPS network. Hyperedges are based on (user, location, activity) relations~\cite{gps5}.
\textit{MovieLens}: a social network where hyperedges are based on (user, movie, tag) relations, describing peoples’ tagging activities~\cite{movielens}.
\textit{Drug}:  a medicine network. The hyperedges are based on (user, drug, reaction) relations~\cite{drug}.\textit{Wordnet}: a semantic network from WordNet 3.0. The hyperedges are based on (head entity, relation, tail entity), expressing the relationships between words~\cite{Wordnet}. \textit{Friendster}: an on-line gaming network. Users can form a group on Friendster social network which other members can then join. These user-defined groups are considered as communities. Communities larger than 500 were removed~\cite{friendster}.

We used the same training and test data setups as Hyper-SAGNN (except for Friendster, which Hyper-SAGNN could not run). For this task, they randomly hide $20$ percentage of existing hyperedges and use the rest of the hypergraph for training. The negative samples are $5$ times the amount of positive samples. We downloaded their code and datasets from their GitHub repository. We used the encoder-based approach to generate the features.

For HyperNetVec, we use two levels of coarsening and two levels of refinement. We first obtain the embedding of the hypergraphs with node2vec as the initial embedding technique. To train our classifier, we used the same positive samples as Hyper-SAGNN. For negative samples we used only the negative samples of a *single* epoch of Hyper-SAGNN.  We then use the vector of the variances of each dimension of the embedding for hyperedge prediction. The intuition is that if nodes are spread out (high variance in the embedding), then they probably do not form a hyperedge whereas nodes that are close to each other are likely to constitute a hyperedge. Various operators such as average, min, and max can be used instead of variance.

\textbf{Experimental results.} Table~\ref{tb:auc} summarizes the hyperedge prediction results for HyperNetVec, node2vec, and Hyper-SAGNN. HyperNetVec achieves the best AUC and running time compared to Hyper-SAGNN. Hyper-SAGNN took almost a day for wordnet whereas HyperNetVec completed the task in less than a minute. HyperNetVec achieves better AUC compared to node2vec on all datasets except drug, and it is always faster.

\textbf{HyperNetVec for large hypergraphs.} We study the scalability of HyperNetVec on a large hypergraph (Friendster) and compare HyperNetVec's accuracy and running time with that of MILE, and DeepWalk (a random-walk based method ~\cite{deepwalk}). (Hyper-SAGNN, GraphZoom and node2vec failed to generate results for Friendster). 
We randomly hide $20\%$ of existing hyperedges and use the rest of the hypergraph to generate the embeddings for the nodes of the hypergraph and finally, use the variance operator for prediction. Since the hypergraph is large, we used five levels of coarsening and ten levels of refinement. The other baseline that was able to run Friendster was MILE with 15 levels of coarsening, and it failed for smaller numbers of coarsening levels. It took MILE 8 hours to generate embeddings for Friendster with an accuracy of 90.4 while it took HyperNetVec only fifteen minutes to do the same task with better accuracy (92.3\%). DeepWalk was also able to generate the embeddings for Friendster after 17 hours with accuracy 84.9\%.

\begin{figure}[tbh]
\begin{center}
    \begin{subfigure}[b]{0.48\textwidth}
            \includegraphics[width=\textwidth]{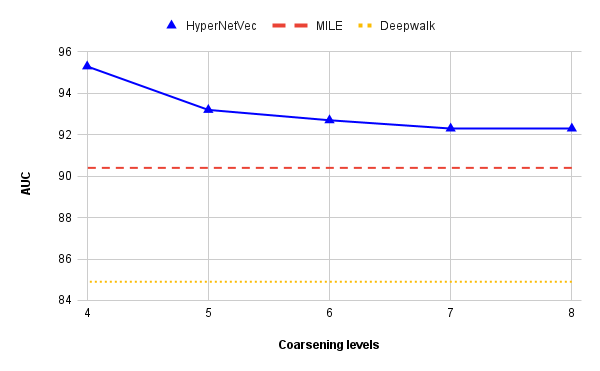}
    \end{subfigure}
    \begin{subfigure}[b]{0.48\textwidth}
      \includegraphics[width=\textwidth]{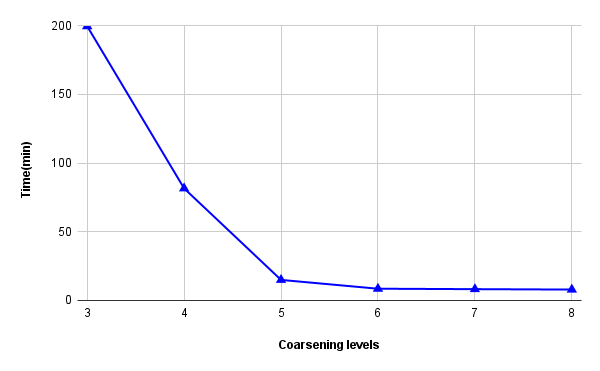}
    
    \end{subfigure}
  \caption{Performance on Friendster } \label{fig:friend}
    \end{center}
    \vskip -0.1in
  \end{figure}

Figure~\ref{fig:friend} compares MILE, DeepWalk, and HyperNetVec in terms of accuracy for different levels of coarsening for HyperNetVec (for MILE, we used 15 levels of coarsening). The main takeaway from Figure~\ref{fig:friend} is that using more levels of coarsening reduces the running time of the overall algorithm, as one would expect. However, a large number of coarsening levels may reduce the accuracy. While this is a fact in most multi-level approaches, Figure~\ref{fig:friend} shows that the loss of accuracy for HyperNetVec is negligible and we are able to get more than 13x speed up by using 5 levels of coarsening instead of 3 levels, while losing less than 3\% in accuracy.

\begin{table*}[htbp]
\caption{Area Under Curve (AUC) scores. Time in seconds.}
\label{tb:auc}
\vskip -0.35in
\begin{center}

\begin{tabular}{lcccccccccc}
\toprule
\textbf{}   &
 \multicolumn{2}{c}{\textbf{GPS}} & \multicolumn{2}{c}{\textbf{MovieLens}} & \multicolumn{2}{c}{\textbf{Drug}} & \multicolumn{2}{c}{\textbf{Wordnet}}& \multicolumn{2}{c}{\textbf{Friendster}} \\ 
  &
  
  \multicolumn{1}{l}{AUC} &
   \multicolumn{1}{l}{Time} &
    \multicolumn{1}{l}{AUC} &
   \multicolumn{1}{l}{Time} &
   \multicolumn{1}{l}{AUC} &
   \multicolumn{1}{l}{Time} &
  \multicolumn{1}{l}{AUC} &
   \multicolumn{1}{l}{Time} &
  \multicolumn{1}{l}{AUC} &
  \multicolumn{1}{l}{Time} \\

\midrule

HyperNetVec   & \textbf{\colorbox{blue!20}{94.5}} & 1   &  \textbf{\colorbox{blue!20}{94.8}}&  6.4  &96.5 & 295   &\textbf{\colorbox{blue!20}{93.0}}&  43.4& \textbf{\colorbox{blue!20}{93.2}} & 897\\
Hyper-SAGNN   & 90.6 & 1,800   &  90.8&  11,160  & 95.9 & 39,540 &87.7& 82,800 & -& - \\
node2vec   & 94.0 &  10  &  79.8 &  19  & \textbf{\colorbox{blue!20}{97.4}}& 895   & 89.0 &  940&-&-\\
\bottomrule
\end{tabular}
\end{center}
\vskip -0.2in

\end{table*}

We also study the behaviour of HyperNetVec for the largest hypergraph, Friendster. In Table~\ref{tb:friends_c}, we see the effect of the coarsening on the size of the hypergraph as well as accuracy and running time. Without coarsening, initial partitioning using DeepWalk takes 17 hours (61,260 seconds) but with 8 levels of coarsening, the end-to-end running time is only 474 seconds. Therefore, using 8 levels of coarsening improves running time by 130X. The table also shows the breakdown of time in different kernels of HyperNetVec. If the coarsest hypergraph is small, most of the time is spent in refinement whereas if the coarsest hypergraph is large, the time is spent mostly in initial embedding. 

\begin{table}[htbp]
\caption{Behavior of HyperNetVec at different levels of coarsening on Friendster. COARSE-N is the number of levels of coarsening, time is in seconds.}
\label{tb:friends_c}
\begin{center}

\begin{tabular}{lccccccc}
\toprule
COARSE-N &0& 3 & 4& 5 & 6 & 7 & 8\\
\midrule
Hyperedges    & 1,616,918&564,262&460,830 & 419,588& 404,857 & 399,194 &396,333\\
Nodes & 7,458,099&436,099&154,418 &85,371& 67,682 &61,669&59,359\\
Coarse Time &0&31&34 &36& 43& 41 & 39 \\
Init  Time &61,260&11,760& 4,620&600& 120& 51 &29\\
Refine Time &0&181&237&261& 261 &349&406\\
Accuracy    &84.9&95.8&95.3& 93.2 &92.7&92.3&92.3\\

\bottomrule
\end{tabular}

\end{center}
\vskip -0.3in
\end{table}

\section{Ongoing work}
\label{sec:ongoing}

In our ongoing research, we want to automatically learn the values of important hyper-parameters such as the number of coarsening levels and the value of $\omega$ that should be used in refinement. For very large hypergraphs, it may be necessary to use distributed-memory machines, which requires partitioning the hypergraph between the memories of the machines in the cluster. Our prior work on BiPart can be used for this task~\cite{bipart,Mateev00,Rogers94}.


%
%
%
\bibliographystyle{splncs04}
\bibliography{mybib}

\begin{thebibliography}{10}
\providecommand{\url}[1]{\texttt{#1}}
\providecommand{\urlprefix}{URL }
\providecommand{\doi}[1]{https://doi.org/#1}

\bibitem{rl}
{Bengio}, Y., {Courville}, A., {Vincent}, P.: Representation learning: A review
  and new perspectives. IEEE Transactions on Pattern Analysis and Machine
  Intelligence  \textbf{35}(8),  1798--1828 (2013). \doi{10.1109/TPAMI.2013.50}

\bibitem{Wordnet}
Bordes, A., Usunier, N., Garcia-Dur\'{a}n, A., Weston, J., Yakhnenko, O.:
  Translating embeddings for modeling multi-relational data. In: Proceedings of
  the 26th International Conference on Neural Information Processing Systems
  (2013)

\bibitem{metis1}
Bui, T.N., Jones, C.: A heuristic for reducing fill-in in sparse matrix
  factorization  (1993)

\bibitem{harp}
Chen, H., Perozzi, B., Hu, Y., Skiena, S.: Harp: Hierarchical representation
  learning for networks (2017)

\bibitem{dblp}
dataset, D.: citation dataset dblp,
  \url{https://www.aminer.org/lab-datasets/citation/DBLP-citation-Jan8.tar.bz}

\bibitem{graphzoom}
Deng, C., Zhao, Z., Wang, Y., Zhang, Z., Feng, Z.: Graphzoom: A multi-level
  spectral approach for accurate and scalable graph embedding (2020)

\bibitem{phmetis}
Devine, K.D., Boman, E.G., Heaphy, R.T., Bisseling, R.H., Catalyurek, U.V.:
  Parallel hypergraph partitioning for scientific computing. In: Proceedings of
  the 20th International Conference on Parallel and Distributed Processing
  (2006), \url{http://dl.acm.org/citation.cfm?id=1898953.1899056}

\bibitem{drug}
FAERS: drug dataset, \url{https://www.fda.gov/Drugs/}

\bibitem{hgnn}
Feng, Y., You, H., Zhang, Z., Ji, R., Gao, Y.: Hypergraph neural networks
  (2019)

\bibitem{cora}
Getoor, L.: Cora dataset, \url{https://linqs.soe.ucsc.edu/data}

\bibitem{crum}
Giurgiu, M., Reinhard, J., Brauner, B., Dunger-Kaltenbach, I., Fobo, G.,
  Frishman, G., Montrone, C., Ruepp, A.: {CORUM: the comprehensive resource of
  mammalian protein complexes-2019.} Nucleic Acids Research  (2019)

\bibitem{fastnv}
Grover, A.: high performance implementation of node2vec,
  \url{https://github.com/snap-stanford/snap/tree/master/examples/node2vec}

\bibitem{node2vec}
Grover, A., Leskovec, J.: node2vec: Scalable feature learning for networks
  (2016)

\bibitem{graphsage}
Hamilton, W.L., Ying, R., Leskovec, J.: Inductive representation learning on
  large graphs (2018)

\bibitem{movielens}
Harper, F.M., Konstan, J.A.: The movielens datasets: History and context. ACM
  Trans. Interact. Intell. Syst.  \textbf{5}(4) (dec 2015).
  \doi{10.1145/2827872}, \url{https://doi.org/10.1145/2827872}

\bibitem{hmetis}
Karypis, G., Aggarwal, R., Kumar, V., Shekhar, S.: Multilevel hypergraph
  partitioning: Applications in vlsi domain. IEEE Trans. Very Large Scale
  Integr. Syst.  (1999)

\bibitem{lossy}
Kirkland, S.: {Two-mode networks exhibiting data loss}. Journal of Complex
  Networks  \textbf{6}(2),  297--316 (08 2017). \doi{10.1093/comnet/cnx039},
  \url{https://doi.org/10.1093/comnet/cnx039}

\bibitem{deeper}
Li, Q., Han, Z., Wu, X.M.: Deeper insights into graph convolutional networks
  for semi-supervised learning (2018)

\bibitem{mile}
Liang, J., Gurukar, S., Parthasarathy, S.: Mile: A multi-level framework for
  scalable graph embedding (2020)

\bibitem{bipart}
Maleki, S., Agarwal, U., Burtscher, M., Pingali, K.: Bipart: A parallel and
  deterministic hypergraph partitioner. SIGPLAN Not.  (2021).
  \doi{10.1145/3437801.3441611}

\bibitem{Mateev00}
Mateev, N., Pingali, K., Stodghill, P., Kotlyar, V.: Next-generation generic
  programming and its application to sparse matrix computations. In:
  Proceedings of the 14th International Conference on Supercomputing. p.
  88–99. ICS '00, Association for Computing Machinery, New York, NY, USA
  (2000)

\bibitem{deepwalk}
Perozzi, B., Al-Rfou, R., Skiena, S.: Deepwalk: Online learning of social
  representations. In: Proceedings of the 20th ACM SIGKDD International
  Conference on Knowledge Discovery and Data Mining. KDD '14, ACM, New York,
  NY, USA (2014)

\bibitem{pingali}
Pingali, K., Nguyen, D., Kulkarni, M., Burtscher, M., Hassaan, M.A., Kaleem,
  R., Lee, T., Lenharth, A., Manevich, R., M{\'{e}}ndez{-}Lojo, M., Prountzos,
  D., Sui, X.: The tao of parallelism in algorithms. In: PLDI 2011. pp. 12--25
  (2011)

\bibitem{disgenet}
Piñero, J., Ramírez-Anguita, J.M., Saüch-Pitarch, J., Ronzano, F., Centeno,
  E., Sanz, F., Furlong, L.I.: {The DisGeNET knowledge platform for disease
  genomics: 2019 update}. Nucleic Acids Research  \textbf{48}(D1),  D845--D855
  (2019)

\bibitem{fact}
Qiu, J., Dong, Y., Ma, H., Li, J., Wang, K., Tang, J.: Network embedding as
  matrix factorization. Proceedings of the Eleventh ACM International
  Conference on Web Search and Data Mining  (2018).
  \doi{10.1145/3159652.3159706}

\bibitem{Rogers94}
Rogers, A., Pingali, K.: Compiling for distributed memory architectures. IEEE
  Transactions on Parallel and Distributed Systems  \textbf{5}(3),  281--298
  (1994)

\bibitem{line}
Tang, J., Qu, M., Wang, M., Zhang, M., Yan, J., Mei, Q.: Line. Proceedings of
  the 24th International Conference on World Wide Web  (2015)

\bibitem{laplacian}
Taubin, G.: A signal processing approach to fair surface design. In:
  Proceedings of the 22nd Annual Conference on Computer Graphics and
  Interactive Techniques (1995). \doi{10.1145/218380.218473}

\bibitem{hypergraph2}
Tu, K., Cui, P., Wang, X., Wang, F., Zhu, W.: Structural deep embedding for
  hyper-networks (2018)

\bibitem{gin}
Xu, K., Hu, W., Leskovec, J., Jegelka, S.: How powerful are graph neural
  networks? (2019)

\bibitem{hypergcn}
Yadati, N., Nimishakavi, M., Yadav, P., Nitin, V., Louis, A., Talukdar, P.:
  Hypergcn: A new method of training graph convolutional networks on
  hypergraphs (2019)

\bibitem{friendster}
Yang, J., Leskovec, J.: Defining and evaluating network communities based on
  ground-truth (2012)

\bibitem{hypergraph3}
Zhang, M., Cui, Z., Jiang, S., Chen, Y.: Beyond link prediction: Predicting
  hyperlinks in adjacency space. In: AAAI. pp. 4430--4437 (2018)

\bibitem{hypersagnn}
Zhang, R., Zou, Y., Ma, J.: Hyper-{SAGNN}: a self-attention based graph neural
  network for hypergraphs. In: International Conference on Learning
  Representations (ICLR) (2020)

\bibitem{gps5}
Zheng, V.W., Cao, B., Zheng, Y., Xie, X., Yang, Q.: Collaborative filtering
  meets mobile recommendation: A user-centered approach. In: Proceedings of the
  Twenty-Fourth AAAI Conference on Artificial Intelligence (2010)

\bibitem{star}
{Zien}, J.Y., {Schlag}, M.D.F., {Chan}, P.K.: Multilevel spectral hypergraph
  partitioning with arbitrary vertex sizes. IEEE Transactions on Computer-Aided
  Design of Integrated Circuits and Systems  (1999). \doi{10.1109/43.784130}

\end{thebibliography}

\end{document}